\documentclass[10pt,twocolumn]{article}

\usepackage[utf8]{inputenc}
\usepackage[T1]{fontenc}

\usepackage{graphicx}

\usepackage{lipsum}
\usepackage{color}

\usepackage{authblk}
\usepackage[margin=2cm]{geometry}
\usepackage{mathtools, cuted}

\usepackage{bm}
\usepackage{amsfonts}
\usepackage{amsmath}
\usepackage{amssymb}
\usepackage{wasysym}
\usepackage{mathrsfs} % cursive letters
\usepackage{siunitx}

\usepackage{hyperref}
\hypersetup{
    colorlinks = true,
    linkcolor = red,
    urlcolor   = black,
    citecolor  = blue
}

\title{Analytical prediction of the temperature and the lifetime of an evaporating spherical droplet}
\author[1]{M. Corpart}
\author[1]{F. Restagno}
\author[1]{F. Boulogne}
\affil[1]{Université Paris-Saclay, CNRS, Laboratoire de Physique des Solides, 91405, Orsay, France.}

\date{\today}
\begin{document}

\twocolumn[
    \begin{@twocolumnfalse}
        \maketitle
        \begin{abstract}
            In this paper, we propose to predict analytically the temperature of an evaporating spherical droplet.
            To do so, we first review, from data in the literature, the effect of temperature on the physical parameters involved in cooling-induced evaporation, namely the saturating vapor pressure, the diffusion coefficient of vapor in air, the liquid density, the enthalpy of vaporization and the thermal conductivity of air.
            These data support a series of approximations that allows us to derive an implicit equation for the liquid temperature.
            We propose a quadratic approximation of the variation of the saturating vapor concentration with temperature to obtain an explicit prediction of the drop temperature.
            As a result, an analytical prediction of the droplet lifetime including the cooling effect is proposed.
        \end{abstract}
    \end{@twocolumnfalse}
]

%%%%%%%%%%%%%%%%%%%%%%%%%%%%%%
%
% INTRODUCTION
%
%%%%%%%%%%%%%%%%%%%%%%%%%%%%%%

\section{Introduction}

Sublimation of solid spheres has been investigated experimentally by Morse in 1910 revealing that the mass loss is not proportional to the surface area but to the radius \cite{Morse1910}.
Langmuir rationalized these findings by considering an adiabatic process where mass transfer is controlled by the diffusion of the vapor in the air~\cite{Langmuir1918}.

The study of spherical droplet evaporation holds significant importance in diverse scientific and technical domains that involve aerosols. Aerosol are produced naturally under different phenomena such as sea spray, fog, clouds, and rain drops. Suspended droplets are also generated by animals and humans during breathing and speaking, which has recently gained attention for airborne contaminants~\cite{Netz2020,Rezaei2021,Pan2022}.
Aerosols can also be produced artificially with spraying techniques for cooling, painting applications, or fuel dispersion in motor engines~\cite{Erbil2012}.
Therefore, understanding the mass transfer of airborne volatile drops is crucial.
This phenomenon is complex due to the coupled heat and mass transfer associated with the phase change, while the transport could occur in a diffusive or a convective manner.
As a result, the theoretical description of the system is more challenging than in the case of the Langmuir adiabatic model.

Therefore, the physics community proposes models to rationalize drop evaporation and predicting their lifetime.
These attempts often involve numerical resolutions of coupled equations that describe transport phenomena, including a broad range of physical effects such as convective or radiative transfer that may occur during the process \cite{Rouault1991,Sobac2015,Andreas1995, Andreas1990, Chini2013, Tonini2012, Benilov2022,Erbil2012}.

At the same time, some studies derived analytical predictions of evaporation kinetics after making several hypothesis \cite{Rezaei2021,Netz2020a, fuchs1960}.
These analytical predictions have the ability to suggest directly how the mechanisms are at play in the quantities of interest.
In particular, the cooling effect due to the enthalpy of vaporization is known to have a significant effect on the drop lifetime \cite{Ranz1952,Ranz1952a,Ranz1956,Beard1971,Pruppacher1979,Andreas1995,Sobac2015,Brutin2015,Schofield2021}.
The variation of temperature in the system leads to the variation of several physical quantities relevant in the process, such as diffusion coefficient, saturating vapor pressure, enthalpy of vaporization, density and thermal conductivity.

In this article, we focus our attention on water spherical droplet evaporating at ambient temperature.
We collect data available in the literature on this system to report the temperature variation of the relevant physical quantities to legitimate upcoming approximations.
Next, we consider the evaporation of the drop in the diffusion-limited regime, and we propose to numerically solve the coupling between evaporation and cooling to obtain the interfacial temperature.
We also use a quadratic approximation to describe the variation of the saturating vapor pressure, which enables to compute analytically the drop temperature and thus its evaporation rate and lifetime.
We compare our description with two other approximations used in the literature, and in particular a linear description of the saturating vapor pressure.
We show that this linear approximation leads to significant differences with the numerical solution, while a quadratic approximation provides an excellent analytical description.

%%%%%%%%%%%%%%%%%%%%%%%%%%%%%%
%
% PHYSICAL PARAMETERS
%
%%%%%%%%%%%%%%%%%%%%%%%%%%%%%%
\section{Temperature variation of some physical constants}

In this section, we present data available in the literature on the temperature variation of the relevant physical constants. Whenever possible, we present experimental data and reference data extracted from a Handbook of chemistry and physics~\cite{Lide2008} and a meteorological table~\cite{List1968}.
We consider the saturating vapor pressure, the diffusion coefficient of vapor in air, the liquid density, the enthalpy of vaporization and the thermal conductivity of air.
We limit our study to an ambient range of temperature between $0$ and $ 30~^\circ$C.

\subsection{Temperature variation of water physical constants}

\begin{figure*}[h!]
    \centering
    \includegraphics[width=.9\linewidth]{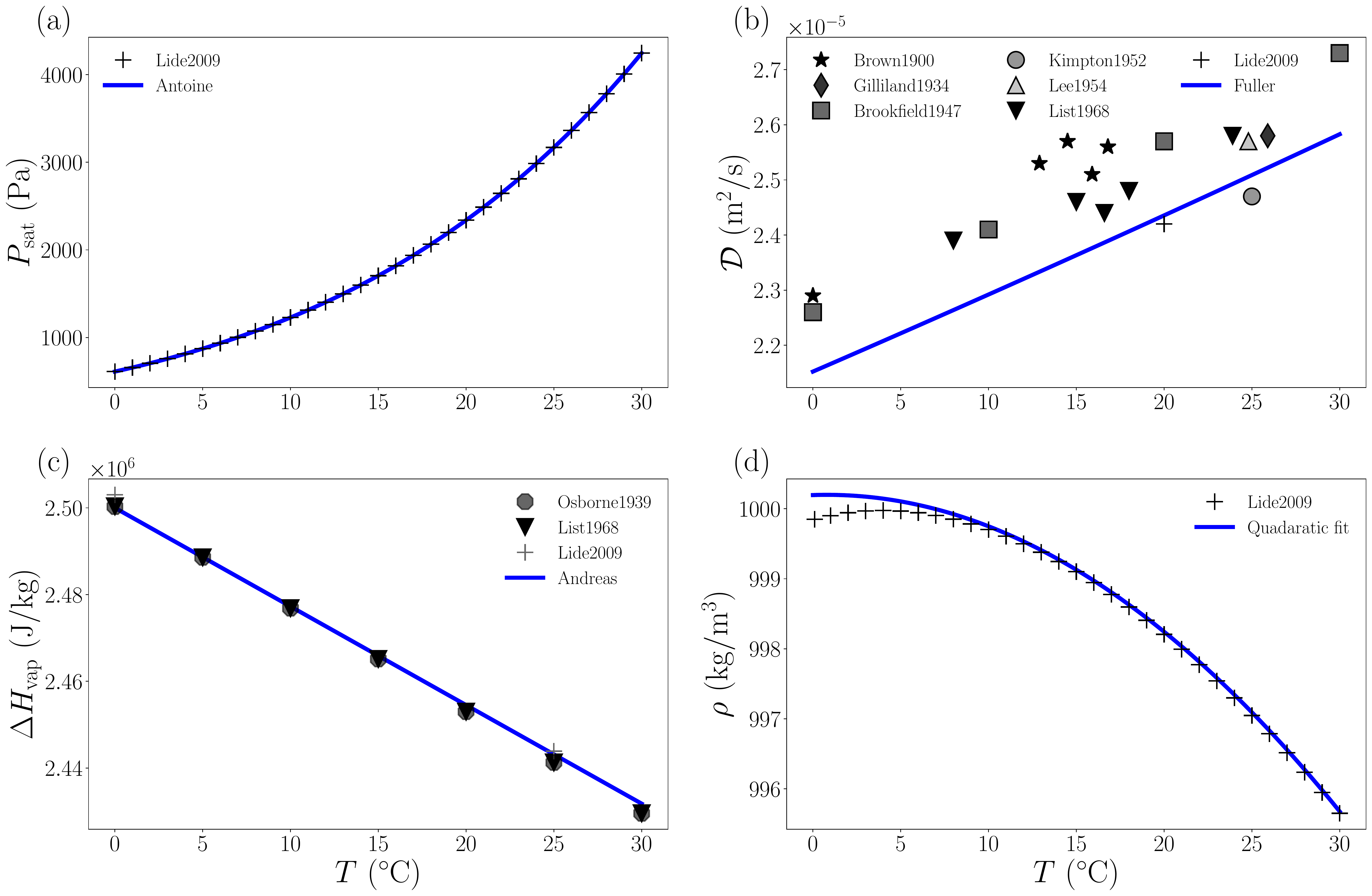}
    \caption{Temperature effect on some physical constants of water. (a) Saturating vapor pressure $P_{\rm sat}$.
    Plus symbols are reference data extracted from a Handbook of chemistry and physics~\cite{Lide2008}, solid blue curve is equation~\eqref{eq:Antoine_equation}.
    (b) Diffusion coefficient ${\cal D}$ of water vapor in air.
    Experimental data are extracted from Brown and Escombe \cite{Brown1900} ($\star$), Gilliland \cite{Gilliland1934} ($\diamond$), Brookfield \textit{et al.} \cite{Brookfield1947} ($\square$), Kimpton and Wall \cite{Kimpton1952} ($\circ$), Lee and Wilke \cite{Lee1954} ($\triangle$).
    Reference data are extracted from List \cite{List1968} ($\triangledown$) and Lide~\cite{Lide2008} ($+$).
    The deep blue curve is Fuller's equation (Eq.~\eqref{eq:Fuller_equation})
    (c) Enthalpy of vaporization of water $\Delta H_{\rm vap}$.
    Experimental data are extracted from Osborne~\cite{Osborne1939} ($\octagon$) and reference data are extracted from List \cite{List1968} ($\triangledown$) and Lide~\cite{Lide2008} ($+$), the deep blue curve is equation~\eqref{eq:h_ev_equation_from_Andreas}.
    (d) Water density $\rho$.
    The reference data are extracted from Lide~\cite{Lide2008} ($+$).
    The deep blue curve is a quadratic fit of the data.
    Fitting parameters are given in equation~\eqref{eq:rho_vs_T_quadratic_fit}.
    }
\label{fig:water_temperature_variation}
\end{figure*}

%%%%%%%%%%%%%%%%%%%%%%%%%%%%%%%%%%%%%
\subsubsection{Saturating vapor pressure $P_{\rm sat}$ and concentration  $c_{\rm sat}$}
%%%%%%%%%%%%%%%%%%%%%%%%%%%%%%%%%%%%%
The measurements of saturating vapor pressure are generally carried out in a closed chamber, containing only the compound to be analyzed, where the temperature is set, and the equilibrium pressure is measured.
The evolution of saturating vapor pressure of water $P_{\rm sat}$ as a function of the temperature $T$ is presented in figure~\ref{fig:water_temperature_variation}(a) where reference data extracted from a Handbook of chemistry and physics~\cite{Lide2008} are symbolized by the plus symbols. It shows a significant increase of the saturating vapor pressure with the temperature.

The variation of the saturating vapor pressure $P_{\rm sat}$ with temperature satisfies the Clausius-Clapeyron~\cite{fuchs1960, Sobac2015} equation
\begin{equation}
    \frac{\mathrm{d} P_{\rm sat}}{\mathrm{d}T}= \frac{\Delta H_{\rm vap} M_{\rm w} P_{\rm sat}}{\mathcal{R} T^2},
\end{equation}
where $\Delta H_{\rm vap}$ is the enthalpy of vaporization of the considered material, here water, $M_{\rm w}$ its molar mass and ${\cal R} \simeq 8.314$ J$\cdot$mol$^{-1}\cdot$K$^{-1}$.
Assuming that the enthalpy of vaporization does not depend on temperature, the Clausius-Clapeyron equation becomes
\begin{equation}\label{eq:equation_Clapeyron}
    \frac{P_{\rm sat}}{p^\circ} = \exp{\left( \frac{- \Delta H_{\rm vap} M_{\rm w}}{{\cal R}}\left(\frac{1}{T} - \frac{1}{T^\circ} \right)\right)},
\end{equation}
where $p^\circ$ and $T^\circ$ are the pressure and temperature of a reference boiling point.

A more robust equation, known as Antoine equation, can be obtained with an additional fitting parameter \cite{Rodgers1978}
\begin{equation} \label{eq:Antoine_equation}
    P_{\rm sat}(T) = p^\circ\, 10^{A -B / (C + T)},
\end{equation}
where $p^\circ = 10^5$~Pa and $A$, $B$, $C$ are constants.
For water at $T \in [0, 30]~^\circ$C, $A, B, C$ are obtained by fitting the data extracted from~\cite{Lide2008}, with $A = 5.341 \pm 0.003$~K, $B = 1807.5 \pm 1.6$~K, and $C = -33.9 \pm 0.1$~K.
Figure~\ref{fig:water_temperature_variation}(a) shows a nice agreement between the data and the model. The typical error between the data and the fit with these parameters is about 0.1 Pa. Alternative expressions are also available in the literature, such as Buck's relation (See for instance \cite{Andreas1995}),  without any noticeable improvements.

The vapor saturating concentration $c_{\rm sat}$ expressed in kg.m$^{-3}$, can be obtained from the ideal gas law
\begin{equation}\label{eq:c_sat_ideal_gas}
    c_{\rm sat}(T) = \frac{P_{\rm sat}(T)  M_{\rm w}}{ {\cal R}  T},
\end{equation}
where $M_{\rm w}=18.02\cdot 10^{-3}$ kg/mol.
%%%%%%%%%%%%%%%%%%%%%%%%%%%%%%%%%%%%%
\subsubsection{Diffusion coefficient  ${\cal D}$ of water vapor in air}
%%%%%%%%%%%%%%%%%%%%%%%%%%%%%%%%%%%%%

To calculate theoretically the diffusion coefficient of a binary system, a molecular theory of diffusion is developed based on collisions of hard sphere in a gas.
To solve the obtained Boltzmann equation the Chapman-Enskog method is used, which gives at first order for a molecule $A$ diffusing in $B$~\cite{Marrero1972}
\begin{equation}
    \mathcal{D}(A,B) =  \frac{8.258 \cdot 10^{-3}}{\sqrt{2}} \frac{\, T^{3/2}\sqrt{\frac{1}{M_{\rm A}} + \frac{1}{M_{\rm B}}}}{P_{\rm atm}\cdot \overline{\Omega}_{A,B}},
\end{equation}
where $M_{\rm A}$ and $M_{\rm A}$ are the molar masses, $T$ is the temperature, $P_{\rm atm}$ is the atmospheric pressure and $\overline{\Omega}_{A,B}$ is the diffusion collision integral for hard spheres.
However, this equation fails to capture the variation of $\mathcal{D}(A,B)$ with the temperature. Indeed, the 3/2 power-law dependence is obtained with an ideal hard spheres model, but experimentally the measured exponent lies between 1.5 and 2 \cite{Fuller1966}. To obtain a more accurate description, additional details on the inter-molecular interactions are required, which depends on the temperature range considered and greatly increases the complexity of the calculation.
That is why most of the time, estimation of diffusion coefficient relies on semi-empirical correlation.
For example, the diffusion coefficient of non-polar gases may be estimated by the Fuller, Schettler, and Giddings' method~\cite{Fuller1966, Fuller1969} that gives
\begin{equation}\label{eq:Fuller_equation}
    \mathcal{D}(A,B) = \frac{ T^{1.75}\sqrt{\frac{1}{M_{\rm A}} + \frac{1}{M_{\rm B}}}}{P_{\rm atm}  \left ( V_A^{1/3} + V_B^{1/3}\right )^2} \cdot 10^{-6}.
\end{equation}
The diffusion coefficient ${\cal D}$ is expressed in m$^2$/s, the molar mass of compound $i$ is in g/mol and $V_i$ is the diffusion volume of the molecule $i$ where $V_i = \sum_{j} n_j V_j$ with $j$ a given atom composing the molecule. The atomic parameters were determined by regression analysis of experimental data and are available in~\cite{Fuller1969}.
For water diffusing in air, the diffusion coefficient is written ${\cal D}$ and we have $V_{\rm air} = 19.7 10^{-6}$~m$^3$/mol, $V_{\rm w} = 13.1 10^{-6}$~m$^3$/mol, $M_{\rm w} = 18.02$~g/mol, and $M_{\rm air} = 28.96$~g/mol~\cite{Fuller1969, Reid1987}.

To test the validity of the Fuller's method, we gathered measurements from different studies \cite{ Brown1900,Gilliland1934,Brookfield1947,Kimpton1952,Lee1954} together with reference values~\cite{List1968, Lide2008} in figure~\ref{fig:water_temperature_variation}(b) and compare them to equation~\eqref{eq:Fuller_equation}.

There are many different ways to measure diffusion coefficient which have various accuracy.
However, to measure diffusion coefficient in air at room temperature, the most used method is the evaporation tube method~\cite{Gilliland1934,Brookfield1947, Kimpton1952,Lee1954}.
Water partially fill a capillary and evaporates into the stagnant gas filling the rest of the tube.
Evaporation rate is measured from the variation of height of the liquid or the variation of weight of the system.
Under the assumptions of quasi-steady evaporation, and, vapor and air being ideal gas, the diffusion coefficient is obtained from the evaporation rate at the temperature of the experiment.
To get precise measurements, the liquid must be carefully kept at constant temperature at each time to avoid evaporative cooling of the liquid.
Indeed, small errors on the estimation of the temperature of the interface and thus on the values of other physical-chemistry parameters (such as $P_{\rm sat}$) can lead to significant errors on the estimated diffusion coefficient.
Finally, surface contamination or convection effects can also lead to inaccurately estimate the diffusion coefficient.
This explains the quite large dispersion of the experimental data in Figure~\ref{fig:water_temperature_variation}.
According to~\cite{Marrero1972}, at best, the reliability of the measurements by the evaporative tube method is several percent ($\approx 10~\%$).
Equation \eqref{eq:Fuller_equation} thus provides a correct estimation of the diffusion coefficient value and its temperature variation, even if it underestimates most of the experimental results plotted in figure~\ref{fig:water_temperature_variation}(b) of about $5~\%$.

Other empirical models~\cite{Andreas1995,Pruppacher1979, Hall1976, fuchs1960} exist to calculate the diffusion coefficient of a $A$ in $B$ at a temperature $T$ but to use them you need to know the value of ${\cal D}(A,B)$ at a given reference temperature and their use does not significantly improve the estimation of the diffusion coefficient.
Moreover, the Fuller's method has the advantage of being easily applicable to the study of other chemical compounds for which experimental measurements of the diffusion coefficient are limited, unreliable or non-existent.

%%%%%%%%%%%%%%%%%%%%%%%%%%%%%%%%%%%%%

%%%%%%%%%%%%%%%%%%%%%%%%%%%%%%%%%%%%%
\subsubsection{Enthalpy of vaporization}
%%%%%%%%%%%%%%%%%%%%%%%%%%%%%%%%%%%%%

Experimental data obtained from calorimetry measurements~\cite{Osborne1939} as well as reference data for enthalpy of vaporization of water $\Delta H_{\rm vap}$ are plotted in figure~\ref{fig:water_temperature_variation}(c)~\cite{List1968, Lide2008}. To predict the evolution of the enthalpy of vaporization of water with temperature, we choose to use the empirical equation given by Fleagle~\cite{Fleagle1981} and Andreas~\cite{Andreas1995}:
\begin{equation}\label{eq:h_ev_equation_from_Andreas}
    \Delta H_{\rm vap} = -2.274 \cdot 10^{3} \, T + 3.121 \cdot 10^{6},
\end{equation}
where $\Delta H_{\rm vap}$  is expressed in J/kg for $T$ in Kelvin~\cite{Fleagle1981, Andreas1995}. Equation~\eqref{eq:h_ev_equation_from_Andreas} is plotted in solid blue curve in figure~\ref{fig:water_temperature_variation}(c). There is a good agreement between both experimental and reference data and equation~\eqref{eq:lambda_air_vs_T_Andreas95}, the difference between equation~\eqref{eq:h_ev_equation_from_Andreas} and the data of the literature being less than $1~\%$.

%%%%%%%%%%%%%%%%%%%%%%%%%%%%%%%%%%%%%
\subsubsection{Liquid density}
%%%%%%%%%%%%%%%%%%%%%%%%%%%%%%%%%%%%%

In figure~\ref{fig:water_temperature_variation}(d), we plot reference values for water density extracted from~\cite{Lide2008} as function of temperature.
We fit the experimental data with a quadratic equation for $T \in [10, 30]^\circ$C and we extend the fit to the entire temperature range which gives
\begin{equation}\label{eq:rho_vs_T_quadratic_fit}
    \rho = -5.3 \cdot 10^{-3} \, T^2 + 2.9 \, T + 6.0 \cdot 10^2,
\end{equation}
where $\rho$ is in kg.m$^{-3}$ and $T$ in Kelvin.
Equation~\eqref{eq:rho_vs_T_quadratic_fit} gives a  good estimation of water density with a an error of the order of $5 \cdot 10^{-2}~\%$ for $T\in [0 ; 10]~^\circ$C and $ 5 \cdot 10^{-3}~\%$ for $T\in [10 ; 30]~^\circ$C.

\subsection{Temperature variation of air thermal conductivity of air}
%%%%%%%%%%%%%%%%%%%%%%%%%%%%%%%%%%%%%

\begin{figure}[h!]
    \centering
    \includegraphics[width=\linewidth]{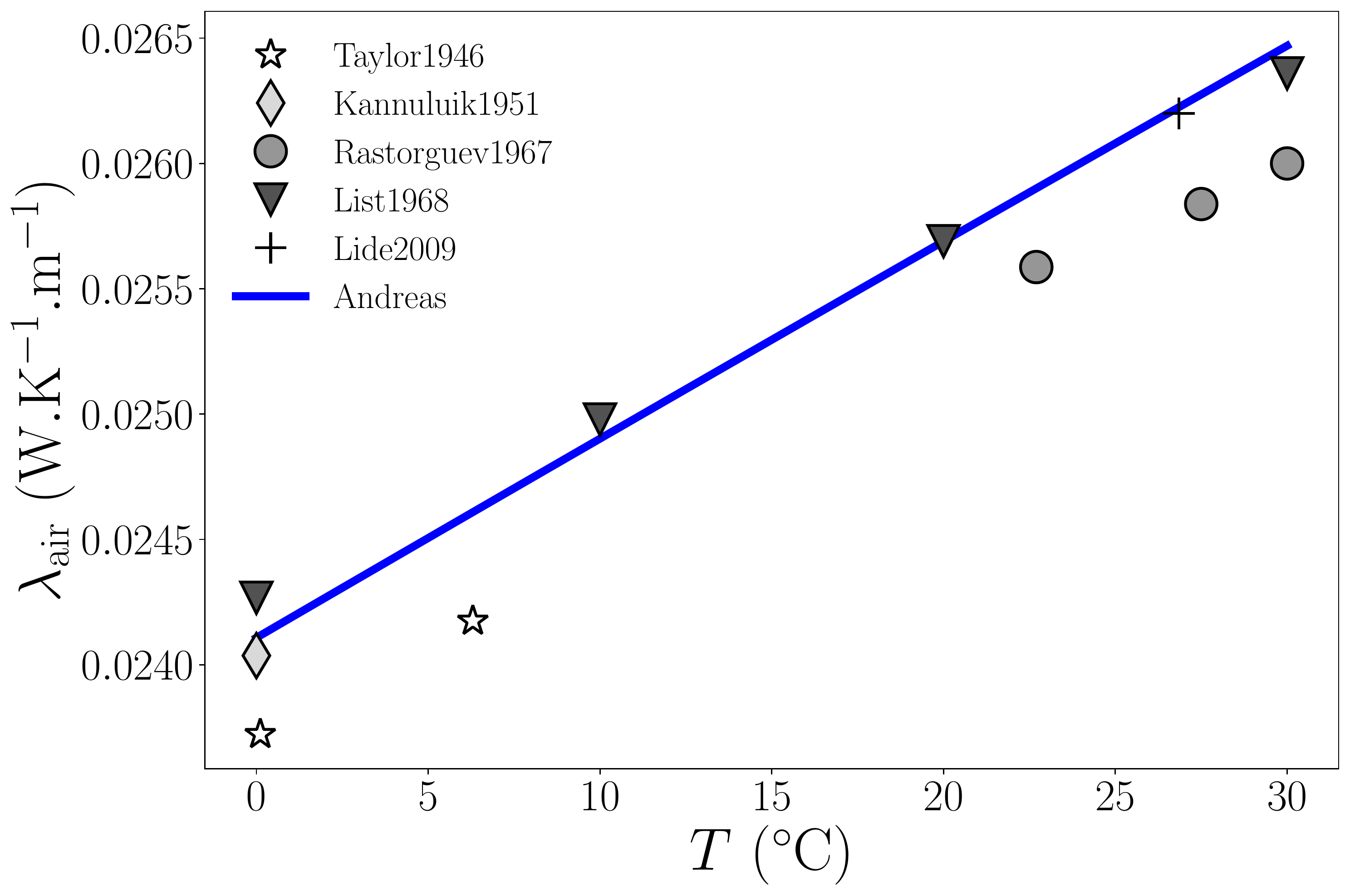}
    \caption{Temperature effect on thermal conductivity of air.
    Experimental data obtained by hot wire measurements extracted from Taylor and Johnston~\cite{Taylor1946} ($\star$), Kannuluik and Carman~\cite{Kannuluik1951} ($\diamond$), Rastorguev and Geller~\cite{Rastorguev1967} ($\circ$) and reference data extracted from List~\cite{List1968} ($\triangledown$) and Lide~\cite{Lide2008} ($+$).
    Deep blue curve is equation~\eqref{eq:lambda_air_vs_T_Andreas95}.}
    \label{fig:air_cond_temperature_variation}
\end{figure}

In figure~\ref{fig:air_cond_temperature_variation}, we plot experimental data of the thermal conductivity of dry air $\lambda_{\rm air}$ measured with the hot wire method~\cite{Taylor1946,Kannuluik1951,Rastorguev1967}. This method consists in recording the temperature variation of a heated wire placed in the fluid of interest to determine its thermal conductivity. We also plot in figure~\ref{fig:air_cond_temperature_variation} the reference data for $\lambda_{\rm air}$ extracted from~\cite{List1968, Lide2008}.

The equation to describe the evolution of the thermal conductivity of dry air with temperature, given by Andreas~\cite{Andreas1995}
\begin{equation}\label{eq:lambda_air_vs_T_Andreas95}
    \lambda_{\rm air}= -3.47 \cdot 10^{-8} \, T^2 + 9.88\cdot 10^{-5} \, T - 2.75\cdot 10^{-4},
\end{equation}
is also represented in figure~\ref{fig:air_cond_temperature_variation}. This equation describes, with a good accuracy, the data from the literature with an error of less than $1~\%$ between equation~\eqref{eq:lambda_air_vs_T_Andreas95} and reference data~\cite{List1968, Lide2008} and an error of $1$ to $2~\%$ with experimental data~\cite{Taylor1946, Kannuluik1951,Rastorguev1967}.
There are other models in the literature to predict the evolution of the thermal conductivity of moist air as a function of temperature and relative humidity ~\cite{Hall1976,Pruppacher1979,tsilingiris2008} but they are more tedious to compute and do not lead to a significant improvement of the description of the data. At $20~^\circ$C, the relative difference between the thermal conductivity of dry and saturated air is about $2~\%$ so we consider that $\lambda_{\rm air}$ is independent of $\mathcal{R}_{\rm H}$ and is equal to the thermal conductivity of dry air. Moreover, the expression given in \cite{Hall1976,Pruppacher1979,tsilingiris2008} for the thermal conductivity of dry and moist air slightly underestimate (error of $1~\% $) the reference data~\cite{List1968, Lide2008}.

\subsection{Summary}
The evolution of the physical parameters of the system with temperature are evaluated with equations~\eqref{eq:Antoine_equation} and \eqref{eq:c_sat_ideal_gas} for $c_{\rm sat}(T)$, \eqref{eq:Fuller_equation} for ${\cal D}$, \eqref{eq:h_ev_equation_from_Andreas} for $\Delta H_{\rm vap}$, \eqref{eq:rho_vs_T_quadratic_fit} for $\rho$ and \eqref{eq:lambda_air_vs_T_Andreas95} for $\lambda_{\rm air}$.
These equations are in good agreement with data from the literature.
From this we can evaluate the variation of all the important parameters when the temperature increases from 0 to 30~$^\circ$C.
This analysis shows that, when $T$ varies from 0 to 30~$^\circ$C, $c_{\rm sat}$ increases by 250$~\%$, ${\cal D}$ increases by 20$~\%$, ${\Delta H_{\rm vap}}$ increases by 3$~\%$, $\rho$ decreases by $0.5~\%$, and $\lambda_{\rm air}$ increases by 8$~\%$.

%%%%%%%%%%%%%%%%%%%%%%%%%%%%%%
%
% MODEL
%
%%%%%%%%%%%%%%%%%%%%%%%%%%%%%%
\section{Model for thermal effect on drop lifetime}

%%%%%%%%%%%%%%%%%%%%%%%%%%%%%%%%%%%%%
\subsection{Equation of mass transfer}
%%%%%%%%%%%%%%%%%%%%%%%%%%%%%%%%%%%%%

We consider the mass transfer of the water vapor in the atmosphere surrounding the spherical drop of radius $R(t)$ and we assume that this process is limited by diffusion, which is valid in a quiescent atmosphere.
This is true for droplet radius significantly larger than the mean-free path of the vapor molecules, \textit{i.e.} $R$ larger than few micrometers~\cite{fuchs1960}.
Over a timescale $R_0^2 / {\cal D}$, where $R_0$ is the initial radius, the transfer can be considered to be in a stationary regime.
In practice, we can check that this timescale is short compared to the total evaporating time, such that the contribution of the starting non-stationary regime is negligible.

Thus, the concentration field $c$ is the solution of the Laplace equation $ \triangle c = 0$, which writes in spherical coordinates
\begin{equation}\label{eq:mass_laplace}
       \frac{1}{r^2}\frac{\mathrm{d}}{\mathrm{d}r}\left(r^2\frac{\mathrm{d}c}{\mathrm{d}r}\right)  = 0.
\end{equation}
This equation is supplemented by two boundary conditions on the concentration, respectively at the liquid-vapor interface and far from the interface,
\begin{align}
    c(r = R) &= c_{\rm sat }(T_{\rm i}), \\
    c(r \to \infty) &= c_\infty,
\end{align}
where $T_{\rm i}$ is the temperature of the interface.
The relative humidity is defined as $\mathcal{R}_{\rm H} = p_\infty/P_{\rm sat}(T_\infty) \approx  c_\infty / c_{\rm sat}(T_\infty)$ in the ideal gas approximation, where $T_\infty$ is the air temperature far from the droplet.

By integrating \eqref{eq:mass_laplace}, the local evaporative flux given by the Fick's law, $j = - {\cal D} \left. \frac{\mathrm{d}c}{\mathrm{d}r}\right|_{r = R}$, writes
\begin{equation}
    j = {\cal D}(T_{\rm i})  \frac{ \Delta c^\star}{R},
\end{equation}
with $\Delta c^\star = c_{\rm sat }(T_{\rm i}) - c_\infty$.

The integration of the local flux over the evaporating surface gives $Q_{\rm ev} = \int j\, {\rm d} S = 4 \pi R \mathcal{D}(T_{\rm i}) \Delta c^\star$, which can be rewritten
\begin{equation}\label{eq:Q_ev_full}
Q_{\rm ev} = 4 \pi R {\cal D}(T_{\rm i}) c_{\rm sat}(T_\infty) \left(\frac{c_{\rm sat}(T_{\rm i})}{c_{\rm sat}(T_\infty)} - \mathcal{R}_{\rm H}\right).
\end{equation}

To compute the evaporation rate $Q_{\rm ev}$ the temperature of the liquid must be determined. To do so, we write in the next paragraph the heat transfer between the atmosphere and the drop.

%%%%%%%%%%%%%%%%%%%%%%%%%%%%%%%%%%%%%
\subsection{Equation of heat transfer}
%%%%%%%%%%%%%%%%%%%%%%%%%%%%%%%%%%%%%

As for the mass transfer, we consider a diffusion limited process in a stationary regime, for which, as for the mass transfer, the air temperature field is a solution of the Laplace equation $ \triangle T = 0$ with the boundary conditions $T(r = R) = T_{\rm i}$ and $T(r \to \infty) = T_\infty$.
The steady-state assumption also implies that the temperature in the drop has reached its equilibrium value $T_{\rm i}$ and is uniform in the liquid.
This is validated if the timescale over which the heat diffuses through the liquid $R_0^2/\kappa{\ell}$  with $\kappa{\ell}$ the thermal conductivity of the liquid, is short compared to the evaporative time~\cite{Sobac2015}.
In practice, this is valid for water droplet evaporating under ambient conditions~\cite{Beard1971, Andreas1995, Sobac2015, Netz2020, Netz2020a}.

The integration of the Laplace equation leads to a total heat flux
\begin{equation}
    Q_{\rm h} = -4 \pi R \lambda_{\rm air}({\overline T}) \Delta T ^\star,
\end{equation}
where $\Delta T ^\star = T_\infty - T_{\rm i}$ and ${\overline T}$ is the average air temperature ${\overline T} = (T_\infty - T_{\rm i})/2$~\cite{Beard1971}.
We assume that the air temperature can be approximated by the effective temperature ${\overline T}$ as done in various studies~\cite{fuchs1960, Erbil2012}.

The heat and mass fluxes are coupled through the enthalpy of vaporization $\Delta H_{\rm vap}(T_{\rm i})$, $ \Delta H_{\rm vap}\,Q_{\rm ev} = - Q_{\rm h}$, which gives
\begin{equation}\label{eq:Delta_T_vs_Delta_c_full}
    T_\infty - T_{\rm i} = \frac{\Delta H_{\rm vap}(T_{\rm i}) {\cal D}\left(T_{\rm i}\right) c_{\rm sat}(T_\infty)}{ \lambda_{\rm air}\left({\overline T}\right)}\left(\frac{c_{\rm sat}(T_{\rm i})}{c_{\rm sat}(T_\infty)} - \mathcal{R}_{\rm H}\right).
\end{equation}

By finding the root of this equation, we can obtain the interface temperature $T_{\rm i}$. We remark that
this temperature is independent of the droplet radius.

%%%%%%%%%%%%%%%%%%%%%%%%%%%%%%
%
% SOLUTIONS
%
%%%%%%%%%%%%%%%%%%%%%%%%%%%%%%
\section{Discussion}

We aim to provide an analytical expression of the interfacial temperature $T_{\rm i}$.
First, we present the results obtained with a numerical approach without further approximation.
Then, we recall approximations found in the literature, and we present a solution based on the quadratic approximation of $c_{\rm sat}(T)$.
All these solutions are compared to the numerical prediction, and we also provide expressions for the drop lifetime.

\begin{figure*}[h!]
    \centering
    \includegraphics[width=.9\linewidth]{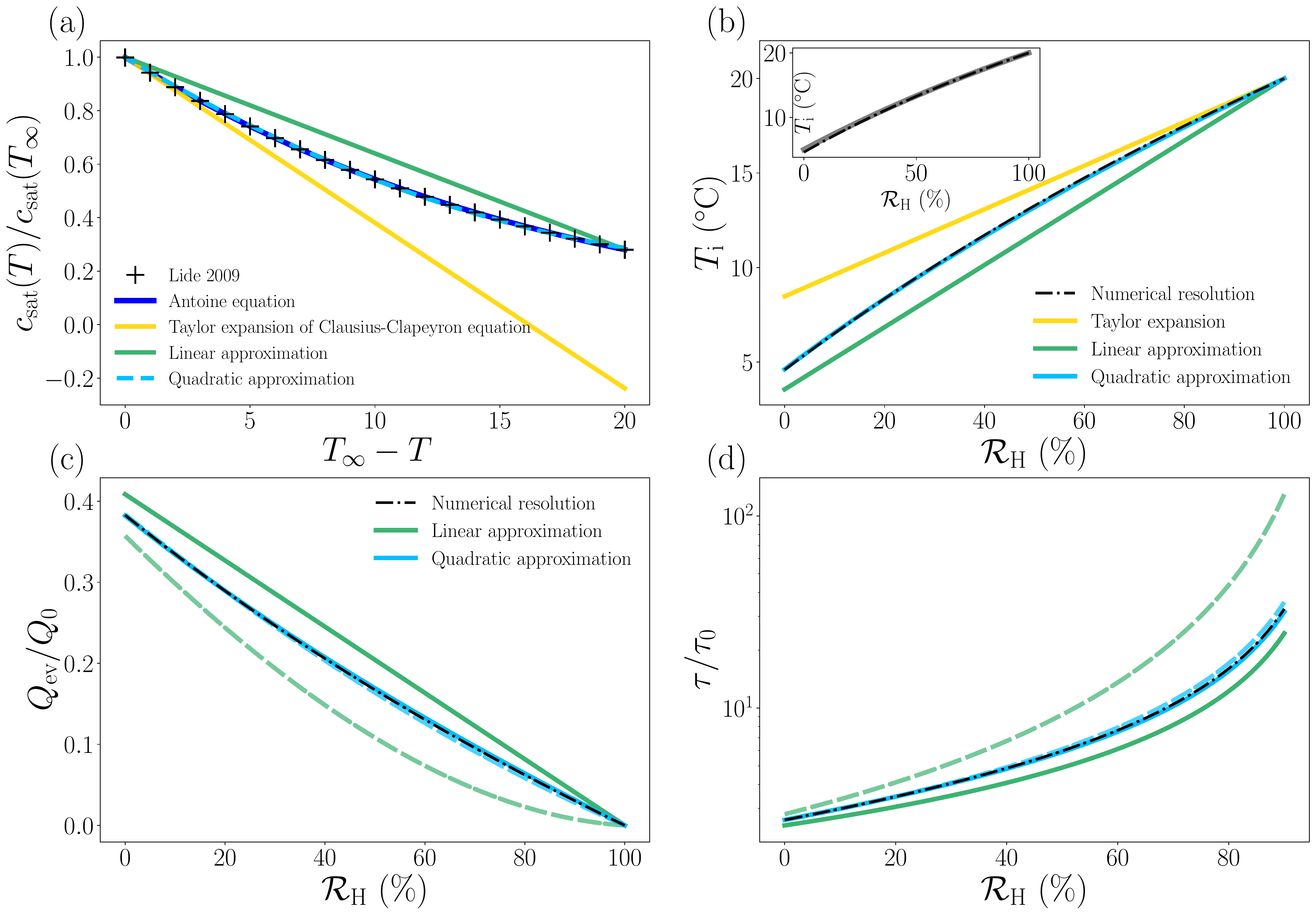}
    \caption{(a) Ratio $c_{\rm sat}(T)/c_{\rm sat}(T_\infty)$ as a function of $T_\infty - T$ for $T_\infty = 20~^\circ$C.
    Black crosses are data from literature presented in Fig.~\ref{fig:water_temperature_variation} combined with equation~\eqref{eq:c_sat_ideal_gas}.
    The deep blue curve is obtained from Antoine equation (Eq.~\eqref{eq:Antoine_equation}).
    The green curve is computed from the linear approximation given by eq.~\eqref{eq:Netz_c_sat_vs_T}.
    The light blue curve is the quadrature defined by equation~\eqref{eq:quadratic_c_sat_vs_T} with the fitting coefficient $\alpha_1 = -5.5\cdot10^{-2} {\rm K}^{-1}$ and $\alpha_2 = 9.8\cdot10^{-4} {\rm K}^{-2}$ at $T_\infty = 20~^\circ$C.
    (b--d) Results obtained for spherical droplets evaporating at $T_\infty = 20~^\circ$C as a function of the relative humidity.
    (b) Interfacial temperature $T_{\rm i}$ obtained with the numerical resolution of equation~\eqref{eq:Delta_T_vs_Delta_c}(dashed-dotted black line), the linear approximation (Eq.~\eqref{eq:Netz_DeltaT}) (green), and the quadratic approximation (Eq.~\eqref{eq:quadratic_Delta_T}) (blue). 
    The inset shows the interfacial temperature obtained by the numerical resolution of equation~\eqref{eq:Delta_T_vs_Delta_c_full} (gray) and of equation~\eqref{eq:Delta_T_vs_Delta_c} (black) as a function of the relative humidity $\mathcal{R}_{\rm H}$.
    Dimensionless (c) evaporative flux $Q_{\rm ev} / Q_0$ and (d) lifetime $\tau/\tau_0$ as a function of $1 - \mathcal{R}_{\rm H}$.
    The numerical resolution is represented in black.
    Results obtained with the linear approximation are computed in green lines and results obtained with the quadratic approximation are represented in light blue lines.
    Solid lines are equations (c) \eqref{eq:Netz_Q_ev} (green) and \eqref{eq:quadratic_Q_ev} (blue) for the evaporative flux and (d) \eqref{eq:Netz_lifetime} (green) and \eqref{eq:quadratic_lifetime} (blue) for the drop lifetime.
    In dashed lines we check the internal coherence of the two approximations by plotting equations (c) \eqref{eq:Q_ev_Antoine_eq_from_T_i} and (d) \eqref{eq:tau_Antoine_eq_from_T_i} in which $T_{\rm i}$ is given by either the linear approximation (eq.~\eqref{eq:Netz_DeltaT}) (dashed green lines) or the quadratic approximation (eq.~\eqref{eq:quadratic_Delta_T} (dashed blue lines).
    }
    \label{fig:all_results}
\end{figure*}

\subsection{Numerical approach}
In this section, we consider the resolution of equation \eqref{eq:Delta_T_vs_Delta_c_full} to obtain the temperature of the interface $T_{\rm i}$ for given atmospheric conditions, namely the temperature $T_\infty$ and the relative humidity $\mathcal{R}_{\rm H}$.
From the interfacial temperature, the concentration ratio $c_{\rm sat}(T_{\rm i}) / c_{\rm sat}(T_\infty)$ can be computed, and thus the drop evaporation and lifetime. The typical evolution of the concentration ratio with the temperature is plotted in figure~\ref{fig:all_results}(a) for $T_\infty = 20~^\circ$C, where the reference data extracted from~\cite{Lide2008} are represented by the $+$ symbols and Antoine equation~\eqref{eq:Antoine_equation} is plotted with a deep blue line.

A numerical approach can be employed to determine the root $T_{\rm i}$ of equation~\eqref{eq:Delta_T_vs_Delta_c_full} by using Newton's method from scipy~\cite{Virtanen2020}, together with equations~\eqref{eq:Antoine_equation}-\eqref{eq:lambda_air_vs_T_Andreas95} to get the temperature evolution of the physical parameters. The temperature of the liquid obtained by the full numerical resolution is plotted in solid gray line in the inset of figure~\ref{fig:all_results}(b).

Nevertheless, the complexity of equation~\eqref{eq:Delta_T_vs_Delta_c_full} prevents analytical solutions. Thus, further approximations must be made to solve equation~\eqref{eq:Delta_T_vs_Delta_c_full}. Due to the weak variations of $\Delta H_{\rm vap}, {\cal D}$ and $\lambda_{\rm air}$ with temperature, we assume that these parameters are independent of the temperature, and we choose to evaluate them at the ambient temperature $T_\infty$.
In this framework, equation~\eqref{eq:Delta_T_vs_Delta_c_full} becomes

\begin{equation}\label{eq:Delta_T_vs_Delta_c}
    T_\infty - T_{\rm i} = \chi \left(\frac{c_{\rm sat}(T_{\rm i})}{c_{\rm sat}(T_\infty)} - \mathcal{R}_{\rm H}\right).
\end{equation}
where $\chi = \Delta H_{\rm vap} {\cal D} c_{\rm sat}(T_\infty) /  \lambda_{\rm air}$.

To test the validity of this hypothesis, we solve numerically (with Newton's method from scipy) equation~\eqref{eq:Delta_T_vs_Delta_c} where $c_{\rm sat}(T)$ is given by Antoine's equation (Eq.~\eqref{eq:Antoine_equation}) under the perfect gas approximation (Eq.~\eqref{eq:c_sat_ideal_gas}).
The temperature of the liquid obtained is plotted in dashed dotted black line in figure~\ref{fig:all_results}(b). Results are in excellent agreement with the full numerical resolution (see the inset of Fig.~\ref{fig:all_results}(b)), the maximum error being about $0.4~^\circ$C for $\mathcal{R}_{\rm H} = 0$.
In the rest of the paper, we will thus work under the assumption that $\Delta H_{\rm vap}, {\cal D}$ and $\lambda_{\rm air}$ are independent of the temperature and their values are taken at $T_\infty$.
Applying this approximation to equation~\eqref{eq:Q_ev_full} we get the evaporation rate of the drop
\begin{equation}\label{eq:Q_ev}
    Q_{\rm ev} = Q_{0} \left(\frac{c_{\rm sat}(T_{\rm i})}{c_{\rm sat}(T_\infty)} - \mathcal{R}_{\rm H}\right)
\end{equation}
with
\begin{equation}\label{eq:Q_0}
    Q_0 = 4 \pi R {\cal D} c_{\rm sat}(T_\infty),
\end{equation}
the evaporation rate of a spherical drop without evaporative cooling ($T_{\rm i} = T_\infty$) and placed in a dry atmosphere ($\mathcal{R}_{\rm H} = 0$).

The droplet lifetime is obtained from the conservation of the drop volume $\Omega = \frac{4}{3} \pi R^3$,
\begin{equation}\label{eq:mass_conservation}
    Q_{\rm ev} = - \rho \frac{{\rm d} \Omega}{{\rm d} t},
\end{equation}
where $\rho$ is the liquid density at the temperature of the liquid $T_{\rm i}$.
We assume that $\rho(T_{\rm i}) = \rho(T_\infty)$, which is fairly reasonable, as $\rho$ decreases by only $0.5~\%$ when the air temperature varies from 0 to 30$^\circ$C.

After integration from $R(0)=R_0$ to $R(\tau)=0$, we have the dynamics of the droplet radius $R(t) = R_0 \sqrt{1 - t / \tau}$, where the droplet lifetime is
\begin{equation}\label{eq:droplet_lifetime}
    \tau =  \tau_0 \left(\frac{c_{\rm sat}(T_{\rm i})}{c_{\rm sat}(T_\infty)} - \mathcal{R}_{\rm H}\right)^{-1}.
\end{equation}
We noted $\tau_0 = \rho R_0^2 / 2 {\cal D} c_\textrm{sat}(T_\infty)$ the lifetime of a spherical drop evaporating in a dry atmosphere ($\mathcal{R}_{\rm H} = 0$) without cooling effect.

Results for evaporation rate and drop lifetime calculated with $T_{\rm i}$ obtained by numerical resolution of equation~\eqref{eq:Delta_T_vs_Delta_c} are plotted in dashed-dotted black lines respectively in figure~\ref{fig:all_results}(c) and in figure~\ref{fig:all_results}(d) as a function of the relative humidity.

The complexity of Antoine equation still prevents us to solve analytically equation~\eqref{eq:Delta_T_vs_Delta_c}. Therefore, we need to establish further approximations on $c_{\rm sat}(T)$ to pursue analytical calculations.
Next, we consider two approximations of the literature, namely a Taylor expansion of Clausius-Clapeyron in the limit $\Delta T \to 0$~\cite{fuchs1960} and a linearized approximation of the evolution of the saturating vapor concentration with the temperature~\cite{Netz2020,Netz2020a}. We also propose to use a quadratic approximation of the variation of $c_{\rm sat}(T)$ and we discuss the level of accuracy of each approach.

\subsection{Taylor expansion of the Clausius-Clapeyron equation}

In the case $T_{\rm i}\approx T_\infty$ and $(T_\infty - T_{\rm i})/T_\infty \ll 1$, Fuchs~\cite{fuchs1960} provides an analytical solution of equation~\eqref{eq:Delta_T_vs_Delta_c} by performing a Taylor expansion of the Clausius-Clapeyron equation~\eqref{eq:equation_Clapeyron} that gives at first order
\begin{equation}\label{eq:Fuchs_c_sat_vs_T}
    \frac{P_{\rm sat}(T_{\rm i})}{P_{\rm sat}(T_\infty)} \approx \frac{c_{\rm sat}(T_{\rm i})}{c_{\rm sat}(T_\infty)} \approx 1 -  \frac{M_{\rm w} \cdot \Delta H_{\rm vap} }{{\cal R}} \frac{T_\infty - T_{\rm i}}{T_\infty ^2}.
\end{equation}
In figure \ref{fig:all_results}(a), we plot this equation in yellow line for $T_\infty=20~^\circ$C. The Taylor expansion of Clausius-Clapeyron equation provides an excellent approximation at $T \approx T_\infty $ but leads to errors of the order of 30~\% at $T = 10~^\circ$C and 200~\% at $T = 0~^\circ$C in the estimation of $c_{\rm sat}(T)/c_{\rm sat}(T_\infty)$.

The substitution of equation~\eqref{eq:Fuchs_c_sat_vs_T} into equation~\eqref{eq:Delta_T_vs_Delta_c} gives an explicit expression for the temperature of the liquid

\begin{equation}\label{eq:Fuchs_Delta_T}
    T_{\rm i} = T_\infty - \frac{\chi}{1 + \chi \frac{\Delta H_{\rm vap} M_{\rm w}}{ \mathcal{R} T_\infty^2 } } \left (1 - \mathcal{R}_{\rm H}\right).
\end{equation}

This equation is represented in yellow line in figure~\ref{fig:all_results}(b).
The temperature drop in a drop of water is of the order of $10~^\circ$C, which invalidates the assumptions used to perform the limited expansion of the Clausius-Clapeyron equation and therefore leads to an incorrect estimation of the temperature in the drop.
Equation~\eqref{eq:Fuchs_Delta_T} underestimates the interface temperature by about $4~^\circ$C for $\mathcal{R}_{\rm H} = 0$.
Thus, this method only provides an analytical prediction of the cooling effect in the vicinity of $\Delta T \to 0$.
Nevertheless, for water, the cooling effect can be significant such that we seek for a more robust prediction.

%%%%%%%%%%%%%%%%%%%%%%%%%%%%%%%%%%%%%
\subsection{Linear approximation of $c_{\rm sat}(T)$}
%%%%%%%%%%%%%%%%%%%%%%%%%%%%%%%%%%%%%

In \cite{Netz2020,Netz2020a}, Netz and Eaton suggest using a linear approximation of $c_{\rm sat}(T)$ to perform an analytical resolution of equation~\eqref{eq:Delta_T_vs_Delta_c}.
We reproduce here this valuable approach and comment it afterwards.
The linearized concentration writes
\begin{equation}\label{eq:Netz_c_sat_vs_T}
    c_\textrm{sat}(T) = c_\textrm{sat}(T_\infty)\left[1 - \Gamma  (T_\infty - T) \right] ,
\end{equation}
where
\begin{equation}\label{eq:Netz_Gamma}
    \Gamma = \frac{1}{\left(T_\infty - T_{\rm m} \right)} \frac{c_\textrm{sat}(T_\infty) - c_\textrm{sat}(T_{\rm m})}{  c_\textrm{sat}(T_\infty)},
\end{equation}
with $T_{\rm m}$ is the melting temperature of the liquid.
In figure \ref{fig:all_results}(a), we plot this equation in green line for $T_\infty=20~^\circ$C. The linear approximation gives a good description of the saturation concentration close to $T_{\rm m}$ and $T_\infty$ but lead to errors higher than $10~\%$ for $T \in [2 ; 15.5]~^\circ$C in the estimation of $c_{\rm sat}(T)/c_{\rm sat}(T_\infty)$  and a maximal error of the order of $20~\%$ for $T \approx 8^\circ$C. As shown in figure~\ref{fig:all_results}(b), the interfacial temperature of the droplet is in the range $[4.5; 15]~^\circ$C for $\mathcal{R}_{\rm H} \in [0; 60]~\%$, the overestimation of $c_{\rm sat}(T_{\rm i})$ will lead to errors when calculating the evaporative flux and the drop lifetime in this humidity range.

Combining equations \eqref{eq:Delta_T_vs_Delta_c} and \eqref{eq:Netz_c_sat_vs_T} leads to an analytical prediction of the temperature difference
\begin{equation}\label{eq:Netz_DeltaT}
    T_\infty - T_{\rm i} = \frac{\chi}{1 + \chi \Gamma} \: \left(1 - \mathcal{R}_{\rm H} \right).
\end{equation}

In figure~\ref{fig:all_results}(b), we plot the interfacial temperature as a function of the relative humidity that we compare to the numerical solution at $T_\infty = 20~^\circ$C.
We observe that in the range $[0,80]~\%$ of relative humidity, the linear approximation underestimate the cooling effect by about $1.5~^\circ$C.

Next, to compute the concentration ratio $c_{\rm sat}(T_{\rm i}) / c_{\rm sat}(T_\infty)$, and therefore the evaporative flux $Q_{\rm ev}$ and the drop lifetime $\tau$, two approaches can be considered.

The first approach consists in keeping the linear approximation (Eq.~\eqref{eq:Netz_c_sat_vs_T}) for using it into equation~\eqref{eq:Q_ev}, which leads to the evaporation rate
\begin{equation} \label{eq:Netz_Q_ev}
  Q_{\rm ev} =   Q_0 \frac{1-\mathcal{R}_{\rm H}}{1 + \chi \Gamma},
\end{equation}
where $Q_0$ is given by equation~\eqref{eq:Q_0}.
In figure~\ref{fig:all_results}(c), the dimensionless evaporative flux $Q_{\rm ev}/Q_0$ is plotted against the relative humidity, and we compare again the linear approximation given by equation~\eqref{eq:Netz_Q_ev} (solid green line) to the numerical solution (black line).
By substituting equation~\eqref{eq:Netz_Q_ev} in equation~\eqref{eq:mass_conservation}, we obtain the lifetime of the drop
\begin{equation} \label{eq:Netz_lifetime}
    \tau = \tau_0  \frac{ 1 + \chi\Gamma }{ 1 - \mathcal{R}_{\rm H}}.
\end{equation}
Equation~\eqref{eq:Netz_lifetime} is plotted in figure~\ref{fig:all_results}(d) in solid green line.

Figure~\ref{fig:all_results}(c) and (d) (solid green lines) shows that the evaporation rate appears to be overestimated by about $20~\%$ leading to underestimating the drop lifetime by about $15~\%$ over the range $\mathcal{R}_{\rm H} = [0,80]~\%$.

The second approach is to consider Antoine equation (Eq.~\eqref{eq:Antoine_equation}), as a better approximation, after using equation \eqref{eq:Netz_c_sat_vs_T} to calculate $T_{\rm i}$, which is at the expense of the internal coherence.
The evaporative flux and the drop lifetime write, respectively,
\begin{align}\label{eq:Q_ev_Antoine_eq_from_T_i}
  Q_{\rm ev} &=   Q_0 \left(\frac{T_\infty}{T_{\rm i}}10^{B\left( [C+T_\infty]^{-1} - [C+T_{\rm i}]^{-1} \right)} - \mathcal{R}_{\rm H}\right), \\
  \tau &= \tau_0 \left(\frac{T_\infty}{T_{\rm i}}10^{B\left( [C+T_\infty]^{-1} - [C+T_{\rm i}]^{-1} \right)} - \mathcal{R}_{\rm H}\right)^{-1}, \label{eq:tau_Antoine_eq_from_T_i}
\end{align}
with $T_{\rm i}$ given by equation~\eqref{eq:Netz_DeltaT}.
The results are plotted in dashed green lines in figure~\ref{fig:all_results}(c) for the evaporation rate and in figure~\ref{fig:all_results}(d) for the drop lifetime.

Due to the overestimation of the cooling effect observed in figure~\ref{fig:all_results}(b) and the accurate description of the saturated pressure in the second step of the calculation, the evaporative flux is now underestimated by $30~\%$ and the lifetime is overestimated by about $50~\%$ and up to 170~$\%$ at high relative humidity ($\mathcal{R}_{\rm H} = 80~\%$).

As a result, we conclude that the linear approximation used with the two previous approaches predicts the correct trends for the evaporation rate and drop lifetime but leads to significantly badly estimated values.
The two approaches are also inconsistent in their predictions.
Naturally, the difference in these physical quantities depends on the atmospheric temperature and relative humidity.
We limited ourselves to a common situation of $T_\infty=20~^\circ$C.
Indeed, for conditions where the interfacial temperature tends either to the atmospheric temperature, \textit{i.e.} at high relative humidity values, or the melting temperature $T_{\rm}$, the linear approximation will be better.

In the next paragraph, we propose to refine the model of the saturated pressure while allowing analytical calculations.

%%%%%%%%%%%%%%%%%%%%%%%%%%%%%%%%%%%%%
\subsection{Quadratic approximation of $c_{\rm sat}(T)$}
%%%%%%%%%%%%%%%%%%%%%%%%%%%%%%%%%%%%%

We refine the model by introducing a quadratic approximation of $c_{\rm sat}(T)$, defined as
\begin{equation}\label{eq:quadratic_c_sat_vs_T}
    c_\textrm{sat}(T) = c_\textrm{sat}(T_\infty)\left(1 + \alpha_1  (T_\infty - T) + \alpha_2  (T_\infty - T)^2 \right),
\end{equation}
where $\alpha_1$ and $\alpha_2$ are obtained by fitting the data from the literature as shown by the light blue curve in figure~\ref{fig:all_results}(a).
For $T_\infty = 20~^\circ$C we have $\alpha_1 = -5.5\cdot10^{-2} {\rm K}^{-1}$ and $\alpha_2 = 9.8\cdot10^{-4} {\rm K}^{-2}$.
The additional order provides a better description of the saturation concentration as shown in figure~\ref{fig:all_results}(a) and equation~\eqref{eq:quadratic_c_sat_vs_T} is an excellent approximation of Antoine equation.

Combining equations~\eqref{eq:Delta_T_vs_Delta_c} and \eqref{eq:quadratic_c_sat_vs_T}, we get
\begin{equation} \label{eq:eq_polynomiale_Delta_T}
    \chi \alpha_2 \: (T_\infty - T_{\rm i})^{ 2} + \left(\chi \alpha_1 - 1 \right) \:(T_\infty - T_{\rm i}) + \chi \left(1 - \mathcal{R}_{\rm H} \right) = 0.
\end{equation}
Among the two roots admitted by equation \eqref{eq:eq_polynomiale_Delta_T}, we keep the one for which $T_\infty - T_{\rm i}$ decreases as $ \mathcal{R}_{\rm H} $ increases, \textit{i.e.}

\begin{equation}\label{eq:quadratic_Delta_T}
    T_\infty - T_{\rm i} = \frac{ 1 -  \chi \alpha_1  - \sqrt{\left(1 - \chi \alpha_1 \right)^2 - 4 \chi^2 \alpha_2 \left(1 - \mathcal{R}_{\rm H} \right)}}{2 \chi \alpha_2}.
\end{equation}
The previous equation is plotted in light blue line in figure~\ref{fig:all_results}(b) and is in excellent agreement  with the numerical solution (black line) of equation~\eqref{eq:Delta_T_vs_Delta_c}.
The quadratic approximation provides a correct description of the liquid temperature over the entire relative humidity range with a maximum deviation of $0.1~^\circ$C.

Then, using again equation \eqref{eq:quadratic_c_sat_vs_T}, the evaporative flux (Eq.~\eqref{eq:Q_ev}) and the drop lifetime (Eq.~\eqref{eq:droplet_lifetime}) can be written

\begin{align}
     \label{eq:quadratic_Q_ev}
    Q_{\rm ev} &= Q_0 \left[\alpha_2 (T_\infty - T_{\rm i})^2  + \alpha_1( T_\infty - T_{\rm i}) + 1 - \mathcal{R}_{\rm H} \right],\\
    \tau &= \tau_0\left[ \alpha_2 (T_\infty - T_{\rm i})^2  + \alpha_1( T_\infty - T_{\rm i})+ 1 - \mathcal{R}_{\rm H} \right]^{-1}, \label{eq:quadratic_lifetime}
\end{align}
where $T_\infty - T_{\rm i}$ is directly provided by equation \eqref{eq:quadratic_Delta_T}.
These equations are plotted in solid blue lines respectively in figures~\ref{fig:all_results}(c) and \ref{fig:all_results}(d), which compare exceptionally well with the numerical resolution and mitigates the error observed with the linear approximation.
Comparison with numerical results shows that quadratic approximation leads to underestimating the drop lifetime of about $1~\%$.
We also checked that, by getting the liquid temperature with equation~\eqref{eq:quadratic_Delta_T} and inserting it into equations~\eqref{eq:Q_ev_Antoine_eq_from_T_i} and \eqref{eq:tau_Antoine_eq_from_T_i}, we get the same results for $Q_{\rm ev}$ and $\tau$ as those obtained with equations~\eqref{eq:quadratic_Q_ev} and \eqref{eq:quadratic_lifetime}.
The two approaches are consistent in their predictions as shown by the superposition of the dashed and solid blue lines in the figure~\ref{fig:all_results}(c) and (d).

%%%%%%%%%%%%%%%%%%%%%%%%%%%%%%
%
% CONCLUSION
%
%%%%%%%%%%%%%%%%%%%%%%%%%%%%%%
\section{Conclusion}
In this paper, we developed an analytical method to predict the lifetime of a spherical drop evaporating in still air by taking into account the evaporative cooling of the liquid.
Here, we focused on water droplets evaporating in still air at ambient temperature, but this study can easily be extended to other liquids and other atmospheric conditions.

First, we used empirical laws to describe the variation of the relevant physical quantities such as diffusion coefficient, saturating vapor pressure, enthalpy of vaporization, liquid density, and air thermal conductivity.
The validity of these empirical laws were testified by comparing their results to the reference data extracted from the literature.
Then, by solving numerically the coupling between mass and heat transfer, we showed that it is sufficient to consider only the variation of the saturating concentration with temperature to describe correctly the evaporative cooling.
To give an analytical description of the diffusion-limited evaporation of a spherical drop, we considered two approximations for the saturated vapor concentration and discussed their validity by comparing their results to the numerical resolution of the problem.
We showed that a linear approximation predicts the correct trends but leads to significant errors on the values of the interfacial temperature, the drop evaporation rate and lifetime.
We thus proposed a quadratic description of the saturating vapor concentration that provides an excellent analytical description of the problem.

\section*{Acknowledgments}
We thank Saint-Gobain and ANRT for funding this study and J. Delavoipière and M. Lamblet for useful discussions.

\bibliography{biblio}

\begin{thebibliography}{10}

\bibitem{Morse1910}
H.~W. Morse.
\newblock On evaporation from the surface of a solid sphere. preliminary note.
\newblock {\em Proceedings of the American Academy of Arts and Sciences},
  45(14):363--367, 1910.

\bibitem{Langmuir1918}
I.~Langmuir.
\newblock The evaporation of small spheres.
\newblock {\em Phys. Rev.}, 12(5):368--370, 1918.

\bibitem{Netz2020}
R.~R. Netz.
\newblock Mechanisms of airborne infection via evaporating and sedimenting
  droplets produced by speaking.
\newblock {\em J. Phys. Chem. B}, 124(33):7093--7101, August 2020.

\bibitem{Rezaei2021}
Majid Rezaei and Roland~R. Netz.
\newblock Airborne virus transmission via respiratory droplets: {Effects} of
  droplet evaporation and sedimentation.
\newblock {\em Current Opinion in Colloid \& Interface Science}, 55:101471,
  2021.

\bibitem{Pan2022}
S.~Pan, C.~Xu, C.~W.~F. Yu, and L.~Liu.
\newblock Characterization and size distribution of initial droplet
  concentration discharged from human breathing and speaking.
\newblock {\em Indoor and Built Environment}, page 1420326X221110975, 2022.

\bibitem{Erbil2012}
H.~Y. Erbil.
\newblock Evaporation of pure liquid sessile and spherical suspended drops: A
  review.
\newblock {\em Adv. Colloid Interface Sci.}, 170(1–2):67 -- 86, 2012.

\bibitem{Rouault1991}
M.~P. Rouault, P.~G. Mestayer, and R.~Schiestel.
\newblock A model of evaporating spray droplet dispersion.
\newblock {\em Journal of Geophysical Research: Oceans}, 96(C4):7181--7200,
  1991.

\bibitem{Sobac2015}
B.~Sobac, P.~Talbot, B.~Haut, A.~Rednikov, and P.~Colinet.
\newblock A comprehensive analysis of the evaporation of a liquid spherical
  drop.
\newblock {\em Journal of Colloid and Interface Science}, 438:306 -- 317, 2015.

\bibitem{Andreas1995}
E.~L. Andreas.
\newblock The temperature of evaporating sea spray droplets.
\newblock {\em J. Atmos. Sci.}, 52(7):852--862, April 1995.

\bibitem{Andreas1990}
E.~L. Andreas.
\newblock Time constants for the evolution of sea spray droplets.
\newblock {\em Tellus B}, 42(5):481--497, November 1990.

\bibitem{Chini2013}
S.~Farshid Chini and A.~Amirfazli.
\newblock Understanding the evaporation of spherical drops in quiescent
  environment.
\newblock {\em Colloids and Surfaces A: Physicochemical and Engineering
  Aspects}, 432:82--88, September 2013.

\bibitem{Tonini2012}
S.~Tonini and G.E. Cossali.
\newblock An analytical model of liquid drop evaporation in gaseous
  environment.
\newblock {\em International Journal of Thermal Sciences}, 57:45--53, July
  2012.

\bibitem{Benilov2022}
E.~S. Benilov.
\newblock Dynamics of a drop floating in vapor of the same fluid.
\newblock {\em Physics of Fluids}, 34(4):042104, April 2022.

\bibitem{Netz2020a}
R.~R. Netz and W.~A. Eaton.
\newblock Physics of virus transmission by speaking droplets.
\newblock {\em Proceedings of the National Academy of Sciences},
  117(41):25209--25211, 2020.

\bibitem{fuchs1960}
N.~A. Fuchs and J.~M. Pratt.
\newblock {\em Evaporation and droplet growth in gaseous media}.
\newblock Pergamon Press Ltd., 1959.

\bibitem{Ranz1952}
W.~E. Ranz and W.R. Marshall.
\newblock Evaporation from drops: Part {I}.
\newblock {\em Chemical Engineering Progress}, 48(3):141--146, 1952.

\bibitem{Ranz1952a}
W.~E. Ranz and W.R. Marshall.
\newblock Evaporation from drops: Part {II}.
\newblock {\em Chemical Engineering Progress}, 48(4):173--181, 1952.

\bibitem{Ranz1956}
W.~E. Ranz.
\newblock On the evaporation of a drop of volatile liquid in high-temperature
  surroundings.
\newblock {\em Trans. of the ASME}, 78(5):909--913, 1956.

\bibitem{Beard1971}
K.~V. Beard and H.~R. Pruppacher.
\newblock A wind tunnel investigation of the rate of evaporation of small water
  drops falling at terminal velocity in air.
\newblock {\em Journal of Atmospheric Sciences}, 28(8):1455 -- 1464, 1971.

\bibitem{Pruppacher1979}
H.~R. Pruppacher and R.~Rasmussen.
\newblock A wind tunnel investigation of the rate of evaporation of large water
  drops falling at terminal velocity in air.
\newblock {\em Journal of the Atmospheric Sciences}, 36(7):1255--1260, 1979.

\bibitem{Brutin2015}
D.~Brutin.
\newblock {\em Droplet Wetting and Evaporation: From Pure to Complex Fluids}.
\newblock Elsevier Science, 2015.

\bibitem{Schofield2021}
F.~G.~H. Schofield, D.~Pritchard, S.~K. Wilson, and K.~Sefiane.
\newblock The lifetimes of evaporating sessile droplets of water can be
  strongly influenced by thermal effects.
\newblock {\em Fluids}, 6(4), 2021.

\bibitem{Lide2008}
D.R. Lide, editor.
\newblock {\em CRC Handbook of Chemistry and Physics}.
\newblock CRC Press/Taylor and Francis, 89th edition edition, 2008.

\bibitem{List1968}
R.~J. List.
\newblock {\em Smithsonian meteorological tables}.
\newblock Smithsonian Institution Press, 6th edition edition, 1968.

\bibitem{Brown1900}
H.~T. Brown and F.~Escombe.
\newblock {VIII.} static diffusion of gases and liquids in relation to the
  assimilation of carbon and translocation in plants.
\newblock {\em Philosophical Transactions of the Royal Society of London.
  Series B, Containing Papers of a Biological Character},
  193(185-193):223--291, 1900.

\bibitem{Gilliland1934}
E.~R. Gilliland.
\newblock Diffusion coefficients in gaseous systems.
\newblock {\em Ind. Eng. Chem.}, 26(6):681--685, June 1934.

\bibitem{Brookfield1947}
K.~J. Brookfield, H.~D.~N. Fitzpatrick, J.~F. Jackson, J.~B. Matthews, E.~A.
  Moelwyn~Hughes, and Alexander~Robertus Todd.
\newblock The escape of molecules from a plane surface into a still atmosphere.
\newblock {\em Proceedings of the Royal Society of London. Series A.
  Mathematical and Physical Sciences}, 190(1020):59--67, 1947.

\bibitem{Kimpton1952}
D.~D. Kimpton and F.~T. Wall.
\newblock Determination of diffusion coefficients from rates of evaporation.
\newblock {\em J. Phys. Chem.}, 56(6):715--717, June 1952.

\bibitem{Lee1954}
C.~Y. Lee and C.~R. Wilke.
\newblock Measurements of vapor diffusion coefficient.
\newblock {\em Ind. Eng. Chem.}, 46(11):2381--2387, November 1954.

\bibitem{Osborne1939}
N.S. Osborne, H.F. Stimson, and D.C. Ginnings.
\newblock Thermal properties of saturated water and steam.
\newblock {\em J. Res. Natl. Bur. Stand., A;()}, 23, 1939.

\bibitem{Rodgers1978}
R.C. Rodgers and G.E. Hill.
\newblock Equations for vapour pressure versus temperature: derivation and use
  of the antoine equation on a hand-held programmable calculator.
\newblock {\em British Journal of Anaesthesia}, 50(5):415--424, 1978.

\bibitem{Marrero1972}
T.~R. Marrero and E.~A. Mason.
\newblock Gaseous diffusion coefficients.
\newblock {\em Journal of Physical and Chemical Reference Data}, 1(1):3--118,
  1972.

\bibitem{Fuller1966}
E.~N. Fuller, P.~D. Schettler, and J.~C. Giddings.
\newblock New method for prediction of binary gas-phase diffusion coefficients.
\newblock {\em Industrial \& Engineering Chemistry}, 58(5):18--27, 1966.

\bibitem{Fuller1969}
E.N. Fuller, K.~Ensley, and J.C. Giddings.
\newblock Diffusion of halogenated hydrocarbons in helium. the effect of
  structure on collision cross sections.
\newblock {\em The Journal of Physical Chemistry}, 73(11):3679--3685, 1969.

\bibitem{Reid1987}
R.C. Reid, J.M. Prausnitz, and B.E. Poling.
\newblock {\em The properties of gases and liquids}.
\newblock McGraw Hill Book Co., New York, NY, 1987.

\bibitem{Hall1976}
W.~D. Hall and H.~R. Pruppacher.
\newblock The survival of ice particles falling from cirrus clouds in
  subsaturated air.
\newblock {\em Journal of Atmospheric Sciences}, 33(10):1995 -- 2006, 1976.

\bibitem{Fleagle1981}
R.~G. Fleagle and J.~A. Businger.
\newblock {\em An introduction to atmospheric physics}.
\newblock Academic Press, 1981.

\bibitem{Taylor1946}
W.J. Taylor and H.L. Johnston.
\newblock An improved hot wire cell for accurate measurements of thermal
  conductivities of gases over a wide temperature range results with air
  between 87 and 375 k.
\newblock {\em The Journal of Chemical Physics}, 14(4):219--233, 1946.

\bibitem{Kannuluik1951}
W.~G. Kannuluik and E.~H. Carman.
\newblock The temperature dependence of the thermal conductivity of air.
\newblock {\em Australian Journal of Chemistry}, 4(3):305--314, 1951.

\bibitem{Rastorguev1967}
Y.L. Rastorguev and V.Z. Geller.
\newblock A modification of the measuring cell for the investigation of the
  thermal conductivity of liquids and gases by the hot-wire method.
\newblock {\em Journal of engineering physics}, 13(1):9--14, 1967.

\bibitem{tsilingiris2008}
P.T. Tsilingiris.
\newblock Thermophysical and transport properties of humid air at temperature
  range between 0 and 100 c.
\newblock {\em Energy Conversion and Management}, 49(5):1098--1110, 2008.

\bibitem{Virtanen2020}
P.~Virtanen, R.~Gommers, T.~E. Oliphant, M.~Haberland, T.~Reddy, D.~Cournapeau,
  E.~Burovski, P.~Peterson, W.~Weckesser, J.~Bright, S.~J. {van der Walt},
  M.~Brett, J.~Wilson, K.~J. Millman, N.~Mayorov, A.~R.~J. Nelson, E.~Jones,
  R.~Kern, E.~Larson, C.~J. Carey, I.~Polat, Y.~Feng, E.~W. Moore,
  J.~{VanderPlas}, D.~Laxalde, J.~Perktold, R.~Cimrman, I.~Henriksen, E.~A.
  Quintero, C.~R. Harris, A.~M. Archibald, A.~H. Ribeiro, F.~Pedregosa, P.~{van
  Mulbregt}, and {SciPy 1.0 Contributors}.
\newblock {{SciPy} 1.0: Fundamental Algorithms for Scientific Computing in
  Python}.
\newblock {\em Nature Methods}, 17:261--272, 2020.

\end{thebibliography}

\bibliographystyle{unsrt}
%\bibliographystyle{abstract} % not nice but helpful to see who's who.

%%%%%%%%%%%%%%%%%%%%%%%%%%%%%%
%
% Misc
%
%%%%%%%%%%%%%%%%%%%%%%%%%%%%%%
\newpage
\clearpage
\section*{Appendix}

\begin{table}[h!]
\centering
\begin{tabular}{cc}
\hline
\multicolumn{1}{|c|}{T~(°C)} &
  \multicolumn{1}{c|}{Psat (Pa)} \\ \hline
\multicolumn{1}{|c|}{0} &
  \multicolumn{1}{c|}{$6,11\times 10^2$}
   \\ \hline
\multicolumn{1}{|c|}{1} &
  \multicolumn{1}{c|}{$6,57\times 10^2$}
   \\ \hline
\multicolumn{1}{|c|}{2} &
  \multicolumn{1}{c|}{$7,06\times 10^2$}
    \\ \hline
\multicolumn{1}{|c|}{3} &
  \multicolumn{1}{c|}{$7,58\times 10^2$}
    \\ \hline
\multicolumn{1}{|c|}{4} &
  \multicolumn{1}{c|}{$8,14\times 10^2$}
    \\ \hline
\multicolumn{1}{|c|}{5} &
  \multicolumn{1}{c|}{$8,73\times 10^2$}
    \\ \hline
\multicolumn{1}{|c|}{6} &
  \multicolumn{1}{c|}{$9,35\times 10^2$}
   \\ \hline
\multicolumn{1}{|c|}{7} &
  \multicolumn{1}{c|}{$1,00\times 10^3$}
    \\ \hline
\multicolumn{1}{|c|}{8} &
  \multicolumn{1}{c|}{$1,07\times 10^3$}
    \\ \hline
\multicolumn{1}{|c|}{9} &
  \multicolumn{1}{c|}{$1,15\times 10^3$}
    \\ \hline
\multicolumn{1}{|c|}{10} &
  \multicolumn{1}{c|}{$1,23\times 10^3$}
    \\ \hline
\multicolumn{1}{|c|}{11} &
  \multicolumn{1}{c|}{$1,31\times 10^3$}
    \\ \hline
\multicolumn{1}{|c|}{12} &
  \multicolumn{1}{c|}{$1,40\times 10^3$}
   \\ \hline
\multicolumn{1}{|c|}{13} &
  \multicolumn{1}{c|}{$1,50\times 10^3$}
   \\ \hline
\multicolumn{1}{|c|}{14} &
  \multicolumn{1}{c|}{$1,60\times 10^3$}
   \\ \hline
\multicolumn{1}{|c|}{15} &
  \multicolumn{1}{c|}{$1,71\times 10^3$}
    \\ \hline
\multicolumn{1}{|c|}{16} &
  \multicolumn{1}{c|}{$1,82\times 10^3$}
    \\ \hline
\multicolumn{1}{|c|}{17} &
  \multicolumn{1}{c|}{$1,94\times 10^3$}
    \\ \hline
\multicolumn{1}{|c|}{18} &
  \multicolumn{1}{c|}{$2,06\times 10^3$}
   \\ \hline
\multicolumn{1}{|c|}{19} &
  \multicolumn{1}{c|}{$2,20\times 10^3$}
    \\ \hline
\multicolumn{1}{|c|}{20} &
  \multicolumn{1}{c|}{$2,34\times 10^3$}
   \\ \hline
\multicolumn{1}{|c|}{21} &
  \multicolumn{1}{c|}{$2,49\times 10^3$}
    \\ \hline
\multicolumn{1}{|c|}{22} &
  \multicolumn{1}{c|}{$2,64\times 10^3$}
    \\ \hline
\multicolumn{1}{|c|}{23} &
  \multicolumn{1}{c|}{$2,81\times 10^3$}
   \\ \hline
\multicolumn{1}{|c|}{24} &
  \multicolumn{1}{c|}{$2,99\times 10^3$}
    \\ \hline
\multicolumn{1}{|c|}{25} &
  \multicolumn{1}{c|}{$3,17\times 10^3$}
    \\ \hline
\multicolumn{1}{|c|}{26} &
  \multicolumn{1}{c|}{$3,36\times 10^3$}
    \\ \hline
\multicolumn{1}{|c|}{27} &
  \multicolumn{1}{c|}{$3,57\times 10^3$}
    \\ \hline
\multicolumn{1}{|c|}{28} &
  \multicolumn{1}{c|}{$3,78\times 10^3$}
    \\ \hline
\multicolumn{1}{|c|}{29} &
  \multicolumn{1}{c|}{$4,01\times 10^3$}
    \\ \hline
\multicolumn{1}{|c|}{30} &
  \multicolumn{1}{c|}{$4,25\times 10^3$}
   \\ \hline
\end{tabular}%
\caption{Saturated vapor pressure for water from \cite{Lide2008}.}
\label{tab:appendix_C_data_psat_vs_T}
\end{table}

\end{document}